\documentclass{eas}
\usepackage{graphicx}
\TitreGlobal{PAH in three dimensional Monte Carlo radiativtive
  transfer}

\begin{document}


\title{PAH in vectorized three dimensional Monte Carlo dust radiative
  transfer models}

\runningtitle{PAH in three dimensional Monte Carlo radiativtive
  transfer}

\runningtitle{PAH in three dimensional Monte Carlo 
  radiativtive transfer} 

\author{Ralf Siebenmorgen}\address{ESO,
  Karl-Schwarzschild-Str. 2, D-85748 Garching b. M\"unchen, Germany}

\author{Frank Heymann}

\address{Astronomisches Institut,
  Ruhr--Universit\"at Bochum, Universit\"atsstra. 150, 44801,
  Bochum, Germany and ESO}

\author{Endrik Kr\"ugel}\address{MPI f\"ur
  Radioastronomie, Auf dem H\"ugel 69, Postfach 2024, D-53010 Bonn,
  Germany}
\begin{abstract}
We present a Monte Carlo (MC) radiative transfer code for complex
three dimensional dust distributions and include transiently heated
PAH.  The correctness of the code is confirmed by comparison with
benchmark results. The method makes use of the parallelization
capabilities of modern vectorized computing units like graphic cards.
The computational speed grows linearly with the number of graphical
processing units (GPU).  On a conventional desktop PC, our code is up
to a factor 100 faster when compared to other MC algorithms. As an
example, we compute the dust emission of proto-planetary disks.  We
simulate how a mid-IR instrument mounted at a future 42m ELT
will detect such disks.  Two cases are distinguished: a homogeneous
disk and a disk with an outward migrating planet, producing a gap and
a spiral density wave.  We find that the resulting mid-IR spectra of
both disks are almost identical.  However, they can be distinguished
at those wavelengths by coronographic, dual-band imaging.  Finally,
the emission of PAHs exposed to different radiation fields is
computed.  We demonstrate that PAH emission depends not only on the
strength but also strongly on the hardness of the radiation, a fact
which has often been neglected in previous models. We find that hard
photons ($>20$\, eV) easily dissociate all PAHs in the disks of T
Tauri stars.  To explain the low, but not negligible detection rate
($<10$\%) of PAHs in T Tau disks, we suggest that turbulent motions
act as a possible path for PAH survival.

\end{abstract}

\maketitle
\section{Dust model}

In the dust model we consider: {\it {large}}
($60\rm{\AA}<a<0.2-0.3\mu$m) silicate (Draine 2003) and amorphous
carbon (Zubko et al. 2004) grains and {\it {small}} graphite
($5<a<80$\,\AA) grains.  For both, we apply a power law size
distribution: $n(a) \propto a^{-3.5}$ and absorption and scattering
cross-sections are computed with Mie theory.  In addition we include
{\it {PAHs}} with 30 and 200 C atoms. For the absorption cross section
(Draine (2010), Malloci (2010), Verstraete (2010)) we follow at photon
energies between 1.7--15\,eV Malloci et al. (2007, their Fig.~4).  At
low frequency, we take a cross section cut-off given as an average of
neutral (Schutte et al. 1993) and ionized (Salama et al. 1996)
species. For hard photons, beyond the 17\,eV band, we scale the
photo-absorption cross section of PAH up to the keV region to similar
sized graphite particles.  By computing cross sections above 100\,eV,
we consider an approximation of kinetic energy losses (Dwek \& Smith
(1996) and apply it to all dust particles. With the advent of
{\it{ISO}} and {\it{Spitzer}} more PAH emission features and more
details of their band structures are detected (Tielens 2008).  We
consider 17 emission bands and take, for simplicity and as suggested
by Boulanger et al. (1998) and Siebenmorgen et al. (1998), Lorentzian
profiles. Parameters are listed in Siebenmorgen \& Kr\"ugel (2001) and
calibrated using mid-IR spectra of starburst nuclei.  For starburst
galaxies, a SED library is computed (Siebenmorgen et al. 2007) and
those models provide good fits to the SED of local galaxies and to PAH
detected at high red shifts ($z \approx 3$, Efstathiou \& Siebenmorgen
2009). In the model, we use dust abundances of [X]/[H] (ppm) of: 31
for [Si], 150 [amorphous C], 50 [graphite] and 30 [PAH], respectively;
which is in agreement with cosmic abundance constrains (Asplund et
al. 2009). We are in the process of upgrading the model to be
consistent with the polarisation of the ISM (Voshchinnikov 2004).

\section{Monte Carlo method}

We compute the radiative transfer with a Monte Carlo (MC) technique,
which allows to handle complicated geometries, by following the flight
path of many random photons.  The basic ideas of our method go back to
Lucy (1999) and Bjorkmann \& Wood (2001).  They are described and
extended to the treatment of transiently heated PAHs in Kr\"ugel
(2006). The space is partitioned into cubes and, where a finer grid is
needed, the cubes are further divided into subcubes (cells).  The star
emits photon packages of constant energy $\varepsilon$, but different
frequencies.  A package entering on its flight path a cell may be
absorbed there or scattered.  The probability for such an interaction
is given by the extinction optical depth along the path within the
cell.  When the package is scattered, it only changes direction
determined in a probabilistic manner by the phase function.  When it
is absorbed, a new package of the same energy, but usually different
frequency $\nu_{\rm new}$ is emitted from the spot of absorption.  The
emission is isotropic.  Each absorption event raises the energy of the
cell by $\varepsilon$, and accordingly its temperature.

\begin{figure} [htb]
\center{\includegraphics[width=7.cm]{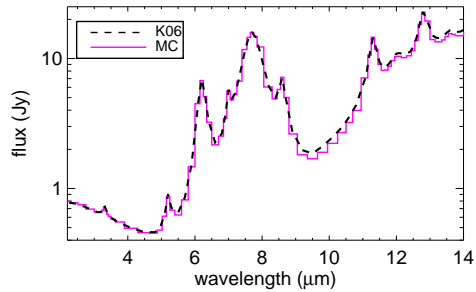}}
\caption{Comparison of a SED of a stellar heated dust sphere of
  constant density and visual extinction to the star of $A_{\rm V}
  =10$\,mag computed with a ray tracing method, as described in
  Kr\"ugel (2006), and the MC treatment of this work; both methods
  agree to within a few \% . \label{sed.ps}}
\end {figure}

We were able to increase the computational speed of the code by up to
two orders of magnitude by parallelizing it, i.e.~by calculating the
flight paths of a hundred or more photons simultaneously.  This allows
the treatment of complex geometries which were so far prohibited
because of their excessive computer demands.  To achieve
parallelization, we had to slightly modify the original code.  The
latter calculates the new frequency $\nu_{\rm new}$ of a package that
is emitted after absorption using the cell temperature before
absorption, but applies a correction for the new temperature due to
Bjorkmann \& Wood (2001).  We now omit this correction or in,
mathematical terms, use Eqn.(11.67) instead of Eqn.(11.65) in Kr\"ugel
(2006). As expected and borne out in tests, the correction may be
neglected when the number of photons absorbed in a cell is not small.
The vectorized MC code was developed during the PhD project of Heymann
(2010).  It was verified against benchmarks provided for stellar
heated dust spheres by Ivezic et al. (1997), the ray tracing code by
Kr\"ugel (2006) and axi--symmetrical disks by Pascucci et al. (2004).
Generally, the computed SED agrees with the benchmarks to within a few
percent.  One example of such comparisons is shown in
Fig.~\ref{sed.ps} for a dust sphere with PAHs. The factor by which
parallelization speeds up the computation scales almost linearly with
the number of graphical processing units.  We point out that
particular attention had to be given to the choice of the random
number generator where we chose the Mersene Twister algorithm
(Matsumoto \& Nishimura, 1998).

\section{Detection of proto-planetary disk structures}

\begin{figure}  [htb]
\hspace{-1.5cm}
\includegraphics[width=8cm]{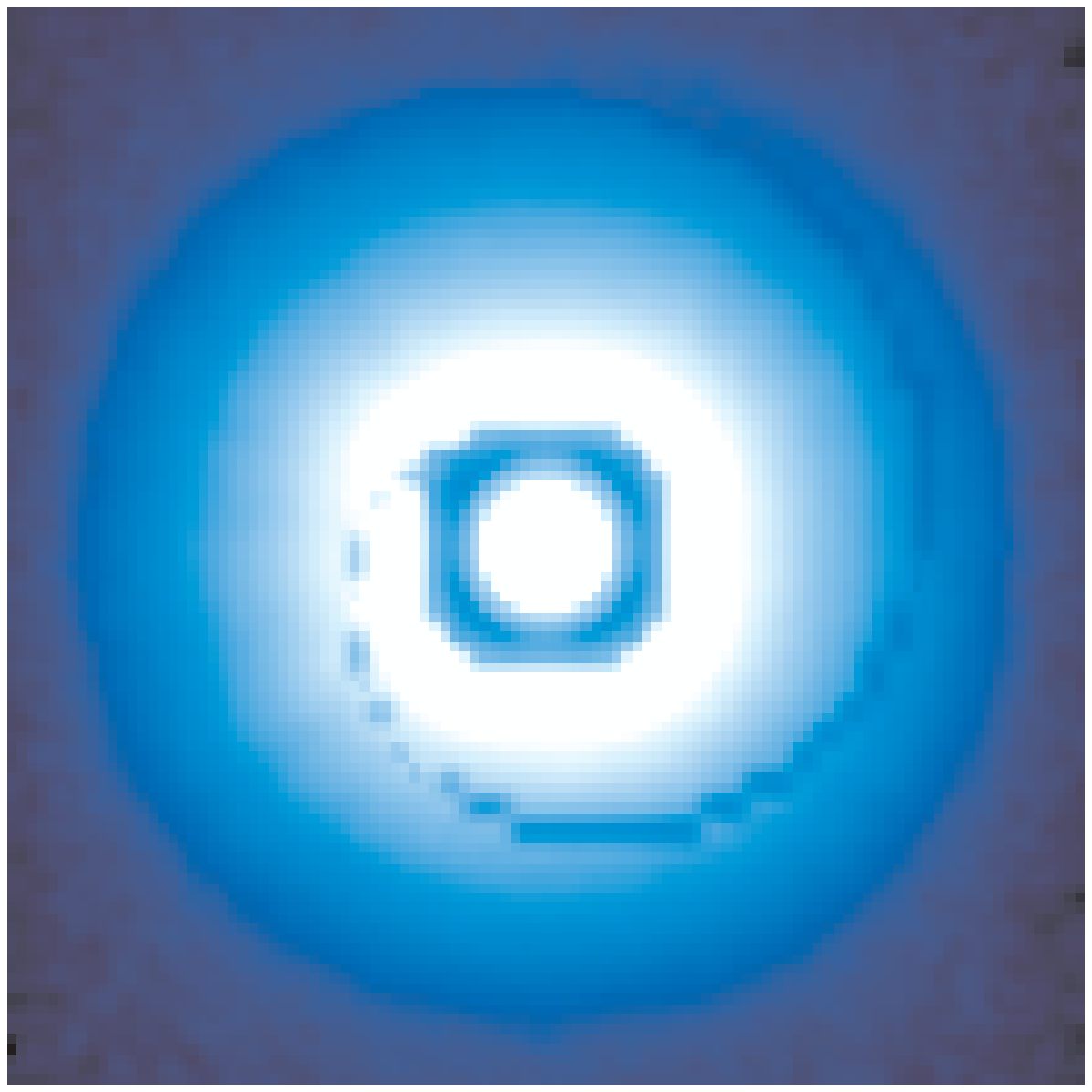}
\hspace{-2cm}
\includegraphics[width=8cm]{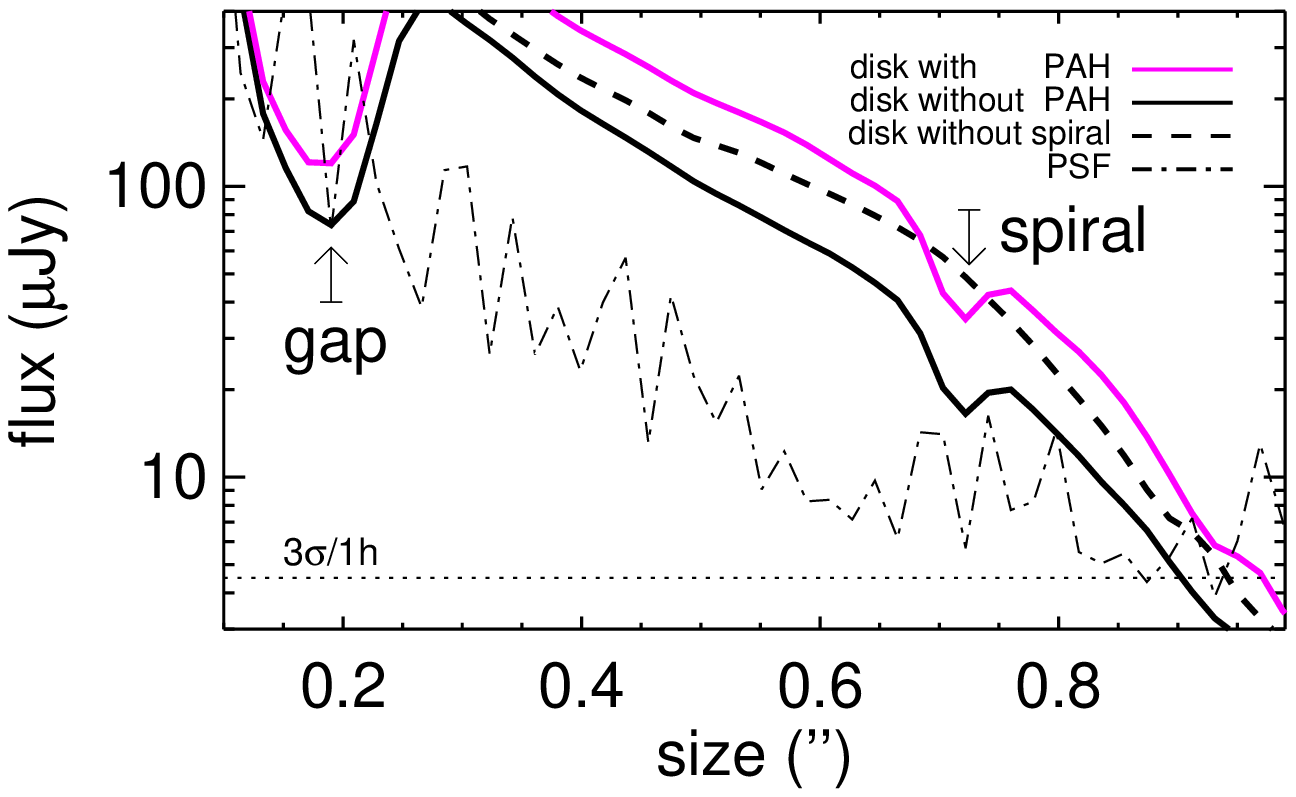}
\caption{Image at 11.3$\mu$m (left) together with flux profiles along
  the major axis (right).  The 3$\sigma$ detection limit (dotted) and
  PSF (dash-dotted) is indicated.  For the disk with (magenta) or
  without (black) PAH the gap and the spiral structure is well
  preserved. A homogeneous disk including PAHs is shown in dased for
  comparison.
\label{metis.ps}}
\end {figure}

Hydrodynamical simulations of proto--planetary disks with an orbiting
planet show particular disk features (Masset et al. 2007). We present
a three dimensional application of the MC code where a disk heated by
a solar type star has a dust density which follows the one as given by
Pascucci et al., and has in addition a gap between 3 -- 8AU and a low
density spiral structure. The optical depth along the midplane from
0.2 to 75\,AU is 10\,mag. We simulate if such disk structures can be
resolved at a distance of 50\,pc with a mid-IR instrument (Brandl et
al. 2010) mounted at a future 42m extreme large telescope (ELT); a
project under study by ESO. The point spread function (PSF) of the ELT
has resolution of 50\,milliarcsec at 10$\mu$m.  In order to improve
the contrast between star and disk, the instrument will provide
coronographic and dual band imaging modes. We choose band passes at
11.3 and 10$\mu$m. The 11.3$\mu$m emission is shown in
Fig.\ref{metis.ps} together with the flux profile along the major
axis. Profiles of such a disk with and without PAHs and that of a
homogeneous disk are computed. The 3$\sigma$ detection limit after 1h
integration is given assuming background limited performance of the
instrument. The models predict that the detailed structure of such a
proto-planetary disk can be well detected.

\begin{figure}  [htb]
\center{\includegraphics[width=7.6cm]{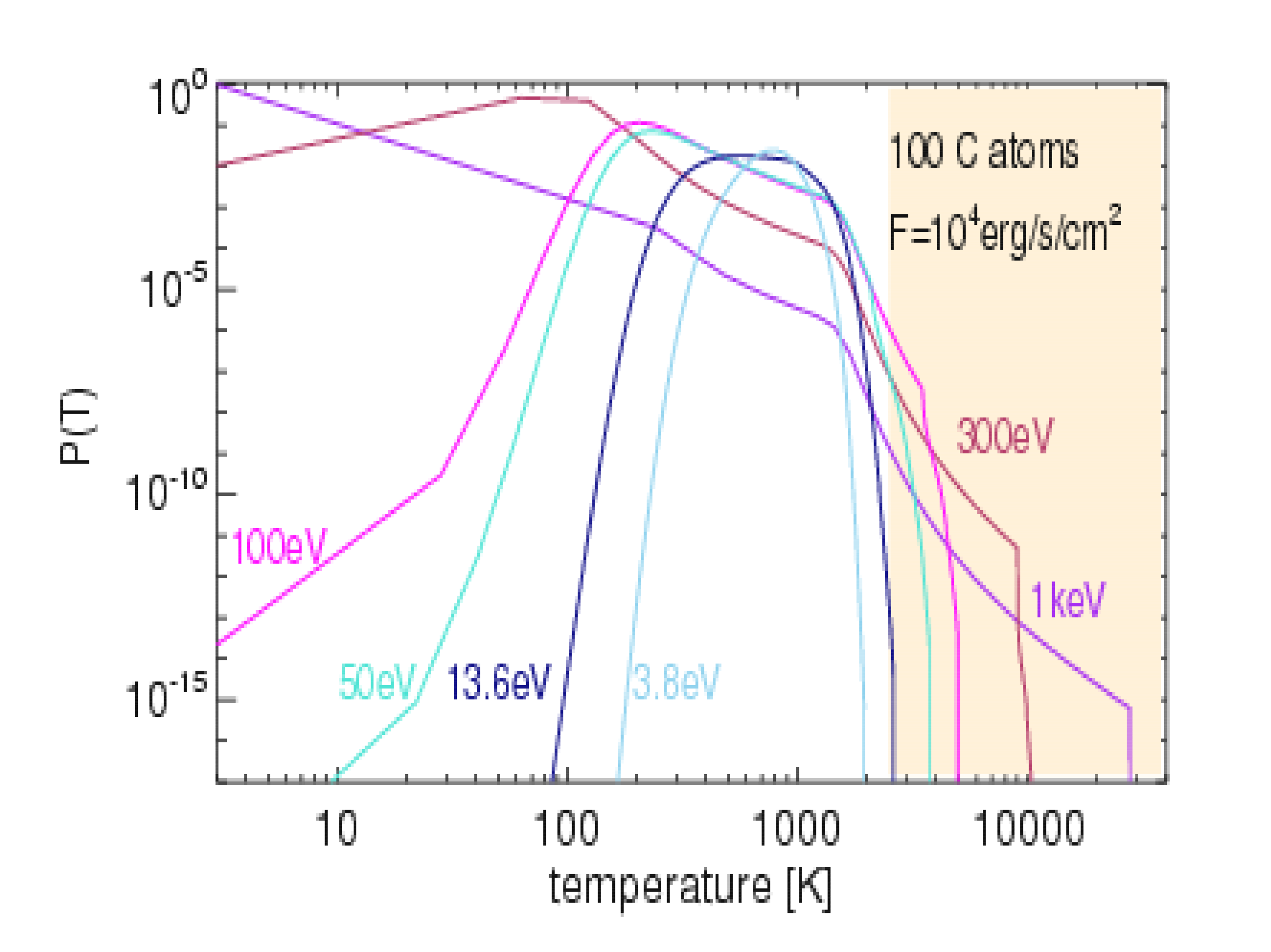}}
\caption{The temperature distribution $P(T)$ of a PAH with 100\,C
  atoms exposed to mono--chromatic radiation with $h\nu= 3.8, 13.6,
  50, 100, 300$\,eV and 1\,keV in a constant heating bath of flux
  $F=10^4$ erg s$^{-1}$ cm$^{-2}$ (Siebenmorgen \& Kr\"ugel 2010).
  The shaded area marks temperatures above the sublimation temperature
  of graphite.
\label{pt.ps}}
\end {figure}

\section{Destruction and survival of PAH in T Tauri disks}

\begin{figure}  [htb]
\center{\includegraphics[width=7cm,angle=270]{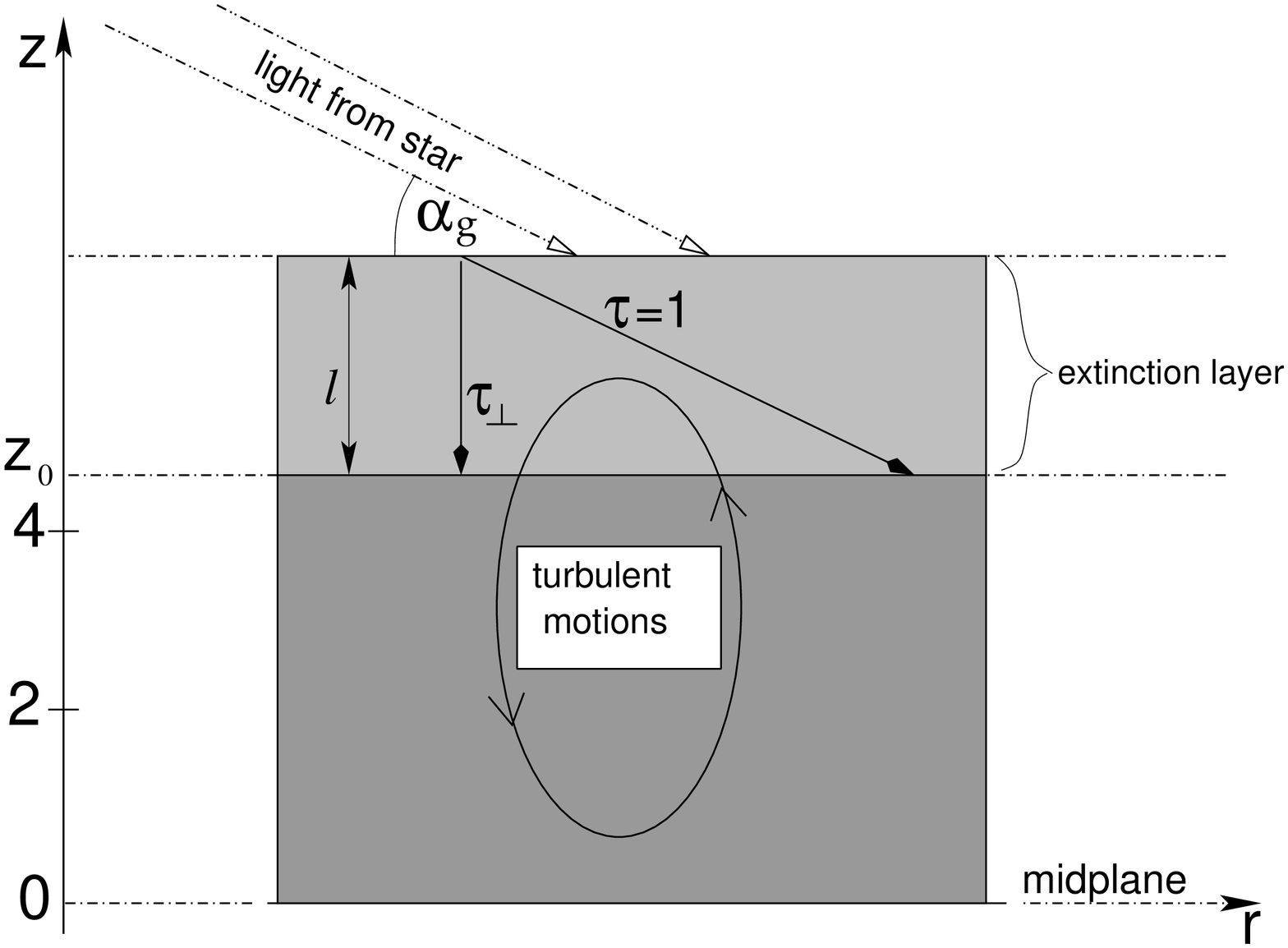}}
\caption{Of each radiation components (photosphere, FUV, EUV and
  X-rays) of the fiducial T Tauri star about $90$\,\% is absorbed in
  what we call extinction layer.  The optical depth from its bottom to
  the star is one and in the vertical direction equal to the grazing
  angle $\alpha_{\rm g}$.  The height of its lower boundary $z_0$
  declines with the radius, but its geometrical thickness is rather
  constant ($\ell \approx 0.5$\,H, for details see Siebenmorgen \&
  Kr\"ugel 2010).  Vertical motions may either remove PAH from the
  extinction layer quickly enough for survival or replenish PAH from
  below.
\label{zdisk.ps}}
\end {figure}

Despite the fact that from the stellar heating one would expect to
detect PAH in the disks of T Tauri, one find them rather seldom, in
less than 10\% of T Tauri (Geers et al., 2006). In order to explain
this fact we present a simplified scheme to estimate the location from
the primary heating source at which the PAH molecules become
photo-stable (Siebenmorgen \& Kr\"ugel 2010).  T Tauri stars have
beside photospheric emission also a far ultraviolet (FUV), an extreme
ultraviolet (EUV) and an X--ray component with a fractional luminosity
of about 1\%, 0.1\% and 0.025\%, respectively.  Such hard photons are
very efficient in dissociating PAHs.  The temperature distribution,
$P(T)$, of PAHs after hard photon absorption is shown in
Fig.~\ref{pt.ps}. It demonstrates that $P(T)$ depends strongly on the
hardness and spectral shape of the exciting radiation field, a fact
which is often neglected in computations of the PAH emission. After
photon absorption, a highly vibrationally excited PAH may relax
through emission of IR photons or, if sufficiently excited, lose atoms
(Omont (1986), and Tielens (2005) for a textbook description). We find
that hard photons (EUV and X-ray) would destroy all PAHs in the disk
of T Tauri stars; whereas soft photons with energies $<20$\,eV
dissociate PAHs only up to short distances from the star (1--2\,
AU). As a possible path for PAH-survival turbulent vertical motions
are suggested. They can replenish or remove PAHs from the reach of
hard photons.  In our treatment the presence of gas is considered
which is ionized at the top of the disk and neutral at lower levels. A
view of the scheme is shown as vertical cut along the midplane in
Fig.~\ref{zdisk.ps}.

\begin{thebibliography}{99}
\bibitem[ ]{ } Asplund, M., Grevesse, N., Sauval, A. J., Scott, P.,
  2009, ARA\&A, 47, 481.

\bibitem[ ]{ } Bjorkman, J.E., Wood, K., 2001, ApJ, 554, 615.

\bibitem[ ]{ } Boulanger, F., Boisssel, P., Cesarsky, D., Ryter, C.,
  1998, A\&A, 339, 194.

\bibitem[ ]{ } Brandl B., \etal\, 2010, SPIE 7735-86, in press.

\bibitem[ ]{ } Draine, B.T., 2003, ApJ, 598, 1026.

\bibitem[ ]{ } Draine, B.T., 2010, this volume.

\bibitem[ ]{ } Dwek, E., Smith, R. K., 1996, ApJ, 459, 686.

\bibitem[ ]{ } Efstathiou, A., Siebenmorgen, R., 2009, A\&A, 502, 541.

\bibitem[]{} Geers V.C., Augereau J.-C., Pontoppidan K. M., et al.,
  A\&A, 459, 545

\bibitem[ ]{ } Heymann, F., 2010, PhD, University of Bochum.

\bibitem[ ]{ } Ivezic, Z., Groenewegen, M. A. T., Men’shchikov, A.,
  Szczerba, R. 1997, MNRAS, 291, 121.

\bibitem[ ]{ } Kr\"ugel E., 2006, An introduction to the Physics of
  Interstellar Dust, IoP.

\bibitem[ ]{ } Lucy, L.B., 1999, A\&A, 344, 282.

\bibitem[ ]{ } Malloci, G., Joblin, C., Mulas, G., 2007, A\& 462, 627

\bibitem[ ]{ } Malloci, G., 2010, this volume.


\bibitem[ ]{ } Masset, F. S., Morbidelli, A., Crida, A., Ferreira, J., 2006, ApJ, 642, 478.

\bibitem[ ]{ } Matsumoto, M.; Nishimura, T., 1998, ACM Transactions on
  Modeling \& Computer Simulation 8 (1): 3–30.

\bibitem[ ]{ } Micelotta, E. R., Jones, A.P., Tielens, A. G. G. M.,
  2010, A\&A, 510, A36.

\bibitem[]{} Omont, A., 1986, A\&A 166, 159.

\bibitem[ ]{ } Pascucci, I., Wolf, S., Steinacker, J., et al. 2004,
  A\&A, 417, 793.

\bibitem[ ]{ } Salama, F., Bakes, E. L. O., Allamandola, L. J., Tielens,
  A. G. G. M. 1996, ApJ, 458, 621.

\bibitem[ ]{ } Schutte, W. A., Tielens A. G. G. M. and Allamandola L. J.,
1993, ApJ 415, 397.


\bibitem[ ]{ } Siebenmorgen, R., Natta, A., Kr\"ugel, E., Prusti, T.,
  1998, A\&A, 339, 134.

\bibitem[ ]{ } Siebenmorgen, R., Kr\"ugel, E., Laureijs, L., 2001, A\&A,
  377, 735.

\bibitem[ ]{ }  Siebenmorgen, R., Kr\"ugel, E., 2007, A\&A, 461, 445. 

\bibitem[ ]{ }  Siebenmorgen, R., Kr\"ugel, E., 2010, A\&A, 511, A6.

\bibitem[ ]{ } Tielens, A. G. G. M., 2005, The Physics and Chemistry
  of the Interstellar Medium, Cambridge Univ. Press.

\bibitem[ ]{ } Tielens, A. G. G. M.,  2008, ARA\&A, 46, 289.

\bibitem[ ]{ } Verstraete, L., 2010, this volume.

\bibitem[ ]{ } Voshchinnikov, N. V., 2004, Optics of cosmic dust
  I, ASPR, 12, 1.

\bibitem[ ]{ }	Zubko, V., Dwek, E., Arendt, R. G.,  2004, ApJS, 152, 211.

\end{thebibliography}
\end{document}